\pdfoutput=1 
\documentclass[aps,pra, amsmath,superscriptaddress, longbibliography,twocolumn, floatfix, 10pt]{revtex4-2}
\usepackage[pdftex]{graphicx}
\usepackage{bm}
\usepackage{setspace}
\usepackage{xr}  %removed since Supplementary moved to Appendix
\usepackage{hyperref}  % uncomment for arxiv 
\usepackage{makecell}
\begin{document} 
\setstretch{1.05}
	
\title{The Role of Defect Geometry in Localized Emission from Monolayer Tungsten Dichalcogenides}

\author{S. Carin Gavin}
\thanks{These authors contributed equally}
\affiliation{Department of Physics and Astronomy, Northwestern University, Evanston, IL 60208, USA}
\affiliation{Pritzker School of Molecular Engineering, University of Chicago, Chicago, IL 60637, USA}
\affiliation{Advanced Institute for Materials Research (WPI-AIMR), Tohoku University, Sendai 980-8577, Japan}
\affiliation{Materials Science Division, Argonne National Laboratory, Lemont, IL 60439, USA}

\author{Moumita Kar}
\thanks{These authors contributed equally}
\affiliation{Department of Chemistry, Northwestern University, Evanston, IL 60208, USA}

\author{Jianguo Wen} 
\affiliation{Center for Nanoscale Materials, Argonne National Laboratory, Lemont, IL 60439, USA}

\author{Anushka Dasgupta}
\affiliation{Department of Materials Science and Engineering, Northwestern University, Evanston, IL 60208, USA}

\author{Jinxuan Pei} 
\affiliation{Department of Physics and Astronomy, Northwestern University, Evanston, IL 60208, USA}

\author{Yiying Liu} 
\affiliation{Department of Physics and Astronomy, Northwestern University, Evanston, IL 60208, USA}

\author{Boyu Zhang} 
\affiliation{Department of Mechanical Engineering, Northwestern University, Evanston, IL 60208, USA}

\author{Charles J. Zeman IV}
\affiliation{Department of Chemistry, Northwestern University, Evanston, IL 60208, USA}

\author{F. Joseph Heremans}
\affiliation{Pritzker School of Molecular Engineering, University of Chicago, Chicago, IL 60637, USA}
\affiliation{Materials Science Division, Argonne National Laboratory, Lemont, IL 60439, USA}

\author{Tobin J. Marks}
\affiliation{Department of Chemistry, Northwestern University, Evanston, IL 60208, USA}
\affiliation{Department of Materials Science and Engineering, Northwestern University, Evanston, IL 60208, USA}
\affiliation{Department of Chemical and Biological Engineering, Northwestern University, Evanston, IL 60208, USA}
\affiliation{Materials Research Center, Northwestern University, Evanston, IL 60208, USA}

\author{Mark C. Hersam}
\affiliation{Department of Chemistry, Northwestern University, Evanston, IL 60208, USA}
\affiliation{Department of Materials Science and Engineering, Northwestern University, Evanston, IL 60208, USA}
\affiliation{Materials Research Center, Northwestern University, Evanston, IL 60208, USA}
\affiliation{Department of Electrical and Computer Engineering, Northwestern University, Evanston, IL 60208, USA}

\author{George C. Schatz} 
\affiliation{Department of Chemistry, Northwestern University, Evanston, IL 60208, USA}

\author{Nathaniel P. Stern}
\email[]{n-stern@northwestern.edu}
\affiliation{Department of Physics and Astronomy, Northwestern University, Evanston, IL 60208, USA}

%\date{\today}
%\email{}

\begin{abstract}

Understanding the mechanism of single photon emission (SPE) in two-dimensional (2D) materials is an unsolved problem important for quantum optical materials and the development of quantum information applications. In 2D transition metal dichalcogenides (TMDs) such as tungsten diselenide ($\text{WSe}_{\text{2}}$), quantum emission has been broadly attributed to exciton localization from atomic point defects, yet the precise microscopic origins are not fully understood. This work introduces an empirically grounded computational framework that explains both the origins of facile SPE in $\text{WSe}_{\text{2}}$ and its relative scarcity in related TMD, tungsten disulfide. High resolution microscopy identifies native defect geometries existing in monolayer $\text{WSe}_{\text{2}}$ lattices providing the ingredients necessary to build a realistic model. The qualitative effects of chalcogen type, defect geometry, and mechanical strain on the electronic structure are then individually assessed using density functional theory, from which a specific divacancy configuration emerges as the candidate for localized single-electron transitions that match observed spectral energies. Spectroscopy and photon correlation measurements further validate this model, establishing a self-consistent link between defect geometry, electronic structure, and quantum emission. By isolating the distinct roles of chalcogen type, defect configuration, and mechanical strain, this work provides a thorough investigation of exciton localization and optical behavior, contributing to a clearer picture of the physical drivers of single photon emission in tungsten-based TMDs.

\end{abstract}
\maketitle

\section{Introduction}

Single photon sources are important to the field of quantum information processing, and single photon emission (SPE) from solid-state materials such as carbon nanotubes, embedded quantum dots, and crystalline defects have collectively made advances as sources of quantum photons \cite{SPEsummary,aharonovich2016solid,shaik2021optical,bathen2021manipulating}. Two-dimensional (2D) materials such as transition metal dichalcogenides (TMDs) are attractive material hosts for SPE given their desirable coupling of optical properties with other degrees of freedom in areas such as valleytronics and spintronics \cite{manzeli20172d,liu2019valleytronics,toriyama2025strategies} and the emerging ability to tailor these properties via heterostructure engineering and surface modification \cite{chemomech,graphite,ananth2025enhanced,dasgupta2025carbene,gavin2025high}. Within this class of materials, monolayer tungsten diselenide ($\text{WSe}_{\text{2}}$) has been especially well-suited to the detection and control of SPE, with extensive work dedicated to utilizing and understanding it \cite{tonndorf,koperski2015single,he2015single,opticalSPE,2013defects,srivastava2015optically,thakar2020optoelectronic,kim2023all,lenferink2022tunable,schwarz2016electrically,clark2016single, palacios2018atomically,graphite,chemomech,dasgupta2025carbene,gavin2025high,wu2025modulation}. The phenomenon of SPE has been broadly attributed to point defects within the 2D lattice that localize excitons \cite{opticalSPE,he2015single,tonndorf,koperski2015single,2013defects}, but despite the abundant experimental observations of SPE in TMD monolayers and heterostructures, its precise microscopic origins remain unclear. This is due in part to technical limitations on measuring the optical properties of single defects in monolayer TMDs as has historically been done in bulk crystals such as diamond \cite{gruber1997scanning}. The defect density in TMDs is high; a typical exfoliated monolayer has an estimated defect density of $10^{12}$ cm$^{-2}$ \cite{defectdensity, aryeetey2020quantification}. A common optical interrogation area such as the diffraction-limited spot size of a laser is on the order of 1 $\mu$m, meaning that optical measurements may excite thousands of defects simultaneously. An observed single photon emitter in the resulting spectra can therefore not be definitively attributed to a specific defect geometry. Furthermore, the variability of SPE phenomena across the TMD family complicates identification of a single underlying mechanism that explains the full picture. Tungsten disulfide ($\text{WS}_{\text{2}}$), molybdenum dilsufide ($\text{MoS}_{\text{2}}$), and molybdenum diselenide ($\text{MoSe}_{\text{2}}$) all share the $\text{MX}_{\text{2}}$ crystal structure and contain similar atomic defects and defect densities \cite{ko2024native}, yet while SPE in $\text{WSe}_{\text{2}}$ is repeatedly observed under the simplest of defect and strain conditions, that from $\text{WS}_{\text{2}}$, $\text{MoS}_{\text{2}}$, and $\text{MoSe}_{\text{2}}$ is relatively scarce and requires intentional and optimized defect and strain engineering techniques to manifest \cite{mose2,uvmos2,ionmos2,chow2015defect,verhagen2020towards,cianci2023spatially}. As such, the origins of SPE in TMDs are inferred indirectly from a combination of microscopy and computation. In previous work, SPE has often been attributed to single chalcogen vacancies due to their relatively low formation energy and identification by scanning tunneling microscopy and scanning transmission electron microscopy (STEM) \cite{liu2013sulfur,ko2024native,zhou2013intrinsic}. Density functional theory (DFT) shows that single vacancies create midgap energy levels in the monolayer electronic structure \cite{irradiation,chemomech,cui2018effect,refaely2018defect,wiktor2016absolute}, but the computational picture is inconsistent between materials and modeling conditions, further obscuring a comprehensive understanding.

Without a definitive understanding of the atomic origins of SPE in TMDs, and defect-bound emission more broadly, this material class cannot be harnessed to its full potential for applications to quantum information. Our work addresses this gap by presenting a thorough computational model supported by microscopy and spectroscopy, focusing on tungsten-based TMDs, which are isolated for comparison here due to their shared transition metal and dark state structure \cite{bao2020probing,godiksen2022impact,kapuscinski2021rydberg,hernandez2022strain}. This isolates the qualitative effect of chalcogen type and defect geometry in our results. High-angle annular dark field STEM (HAADF-STEM) is performed to produce images of defects native to $\text{WSe}_{\text{2}}$ monolayer lattices, providing a realistic base on which to build a computational model. We then present a robust set of DFT calculations that encompass the effect of defect geometry on the electronic structure, proposing a specific divacancy formation as the source of localized single-electron transitions. Finally, photoluminescence (PL) spectroscopy and photon correlation measurements are provided as a self-consistent reference for the spectral properties of the material analyzed in this work. Together, this combination of microscopy, computation, and spectroscopy contributes to a clearer picture of the atomic origins of localized emitters and single photon emission in $\text{WSe}_{\text{2}}$, $\text{WS}_{\text{2}}$, and a broader view of their place among TMDs.

\vfill

\section{Identifying Point Defects}

A critical aspect of this work is understanding defect geometry both experimentally and computationally. Monolayer TMDs, both exfoliated and grown by chemical vapor deposition, are estimated to have typical defect densities of $10^{12} - 10^{13}$ cm$^{-2}$ \cite{defectdensity, aryeetey2020quantification}. This density is assumed to be dominated by single chalcogen vacancies (monovacancies) from the top and bottom atomic planes of the crystal $\text{MX}_{\text{2}}$, indicated by the relatively low formation energy of monovacancies compared to transition metal defects or chalcogen multi-vacancies \cite{liu2013sulfur,ko2024native,zhou2013intrinsic}. Although single chalcogen vacancies may be the dominant point defect, divacancies, consisting of two chalcogen vacancies in nearest neighbor sites, are also prevalent when the defect density is high. Jeong \textit{et al}. suggested that the relative likelihood of monovacancies to hybridize into divacancies is indicated by the single vacancy binding energy  \cite{jeong2019spectroscopic}. In the case where the single vacancy formation energy reveals the energetic cost of vacancy formation in a pristine lattice, the vacancy binding energy describes its thermodynamic stability within the lattice once created. Monovacancies with high binding energies are more likely to form divacancies in the lattice. Among TMDs, $\text{WSe}_{\text{2}}$ has the highest single vacancy binding energy \cite{jeong2019spectroscopic}, providing a variable to consider why SPE reports in this material are so abundant compared to the relative scarcity of reports for $\text{WS}_{\text{2}}$, $\text{MoSe}_{\text{2}}$, and $\text{MoS}_{\text{2}}$ collectively \cite{mose2,uvmos2,ionmos2,chow2015defect,verhagen2020towards,cianci2023spatially}.

\begin{figure}
    \centering
    \includegraphics[width=\linewidth]{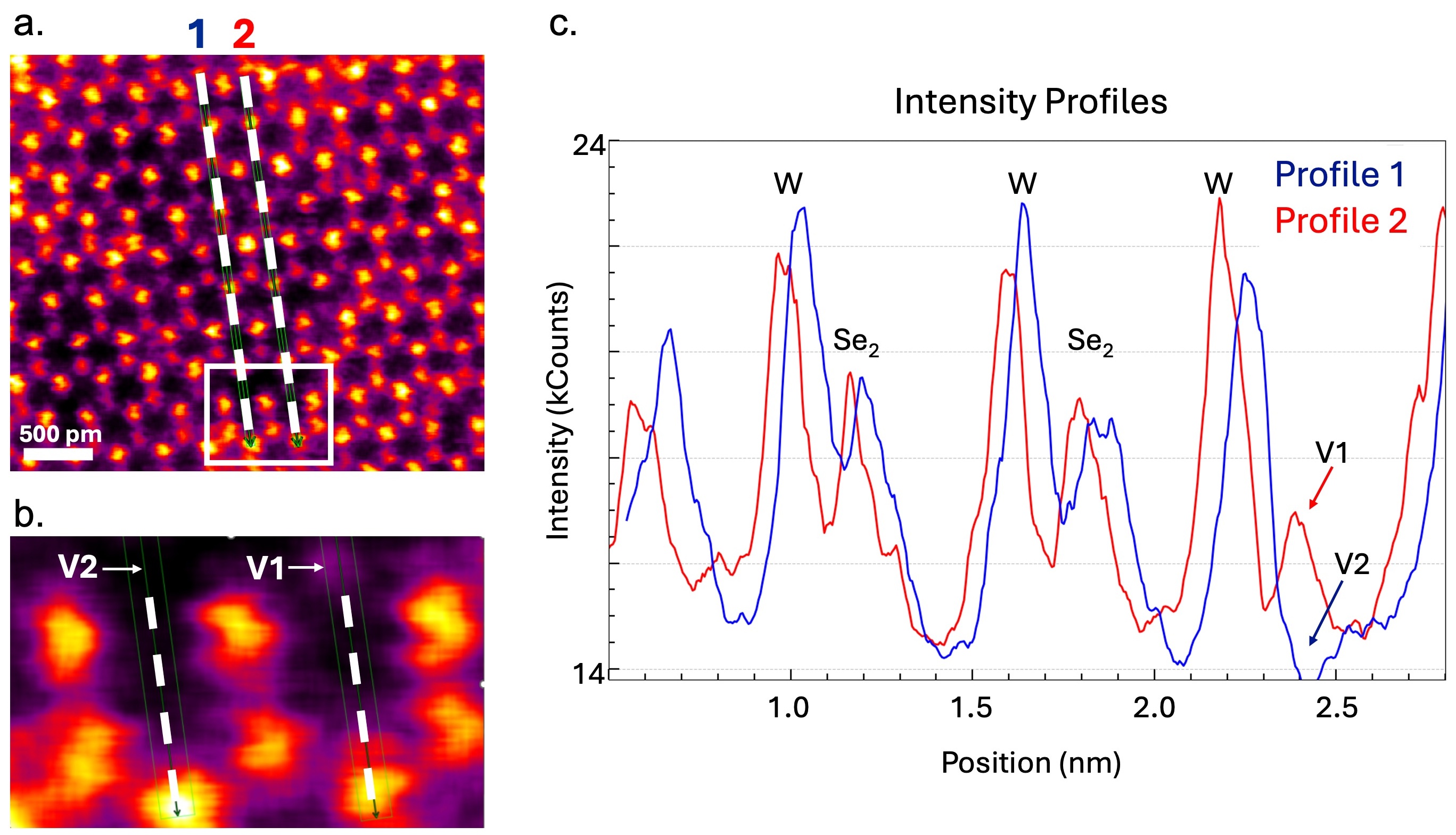}
    \caption{(a) HAADF-STEM image of an approximately 4 nm $\times$ 4 nm area of monolayer $\text{WSe}_{\text{2}}$, with outlined intensity profiles (1,2) quantified in (c). (b) A close-up image of the single and double vacancies outlined by the white box in (a), visually identifying a vertical divacancy (V2) and a single top-plane vacancy (V1).}
    \label{v2v-stem}
\end{figure}

Since SPE in $\text{WSe}_{\text{2}}$ was first observed in regular exfoliated monolayers \cite{tonndorf}, a grounded explanation of its atomic origins starts with experimental demonstration of which defects are native to the monolayer lattice and their relative abundance. Figure \ref{v2v-stem} shows HAADF-STEM images and height profiles of the exfoliated monolayer $\text{WSe}_{\text{2}}$. Figure \ref{v2v-stem}a is a monolayer region of approximately 4 nm $\times$ 4 nm, with line profiles one and two (blue and red, respectively), outlined. In addition to single vacancies, Figure \ref{v2v-stem}b highlights a visual identification of a vertical divacancy (two aligned vacancies from the top and bottom chalcogen planes) compared to a top plane single vacancy, and Figure \ref{v2v-stem}c shows the corresponding line profiles and quantified intensity of a single vacancy (V1) vs. a vertical divacancy (V2). Vertical divacancies in $\text{WSe}_{\text{2}}$ have also previously been identified with STEM, with an observed density of $\sim$ 0.17 $\text{nm}^{\text{-2}}$ \cite{zheng2019point}, supporting their theoretically proposed prominence \cite{jeong2019spectroscopic}. In the 4 nm $\times$ 4 nm region shown in this image, three vertical divacancies are identified, which translates to a density of $\sim$ 0.19 $\text{nm}^{\text{-2}}$, consistent with the previous report. To ensure that the defects observed were native to the exfoliated lattice and not induced by the beam, the scanning beam voltage was restricted to 80~kV, keeping it well below the knock-off energy of chalcogen in the lattice \cite{Zhaoetal2017,Dashetal2021}. Additional STEM images of the V2 geometry are found in Supplemental Information, Figure \ref{supp-1425-big}. Given our STEM imaging and the literature precedents discussed, the V2 geometry is important to investigate in a systematic analysis of defects on exciton localization and SPE.

\begin{figure}[htbp]
	\centering
	\includegraphics[width=\columnwidth]{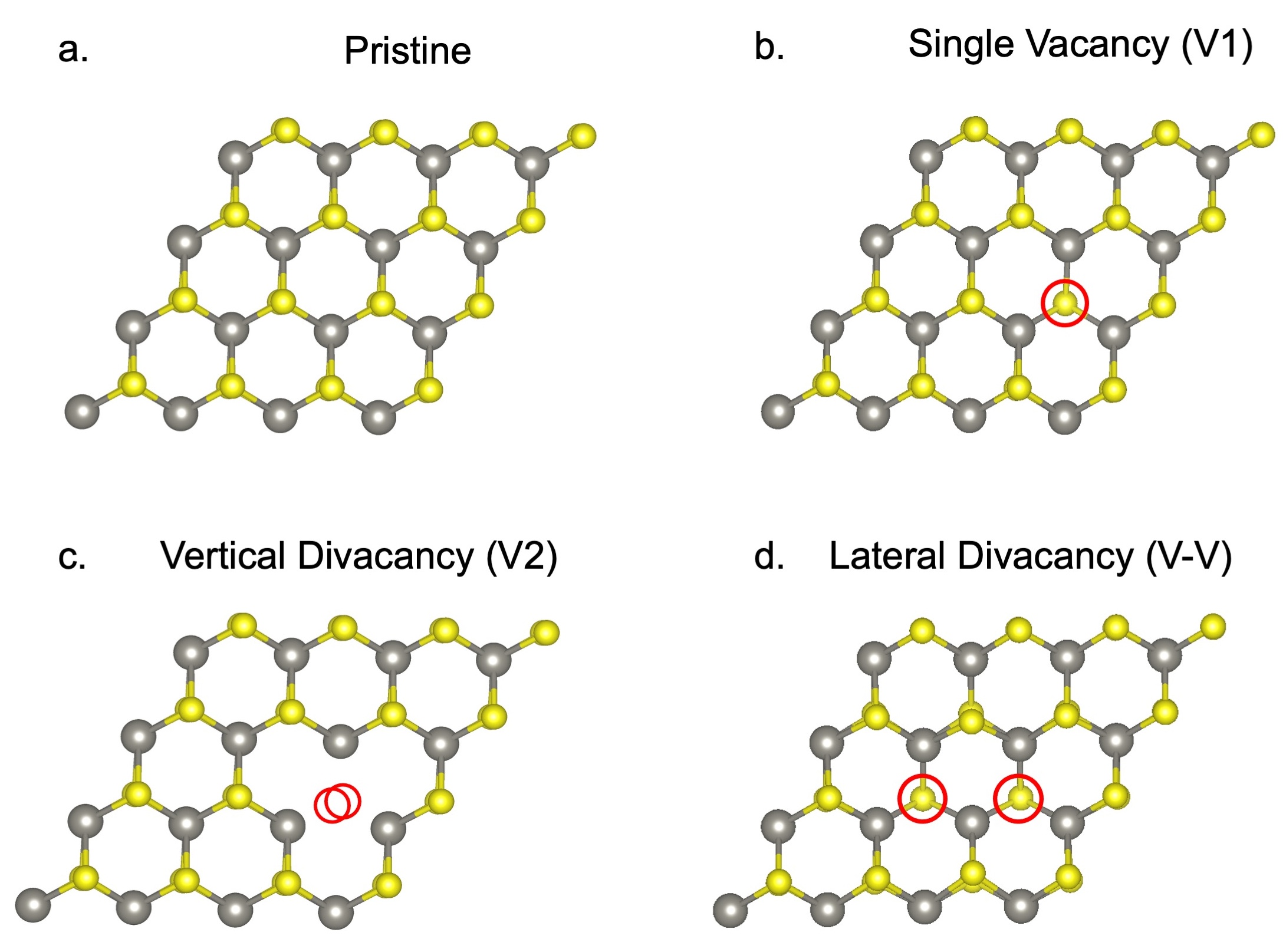}
	\caption{Schematic of the pristine crystal and modeled defect configurations: monovacancy (V1), vertical divacancy (V2), and lateral divacancy (V-V). Chalcogen atoms are shown in yellow, while tungsten atoms are shown in grey.}
	\label{vacancyconfigs}
\end{figure}

\begin{figure*}[tbph]
	\centering
	\includegraphics[width=\textwidth]{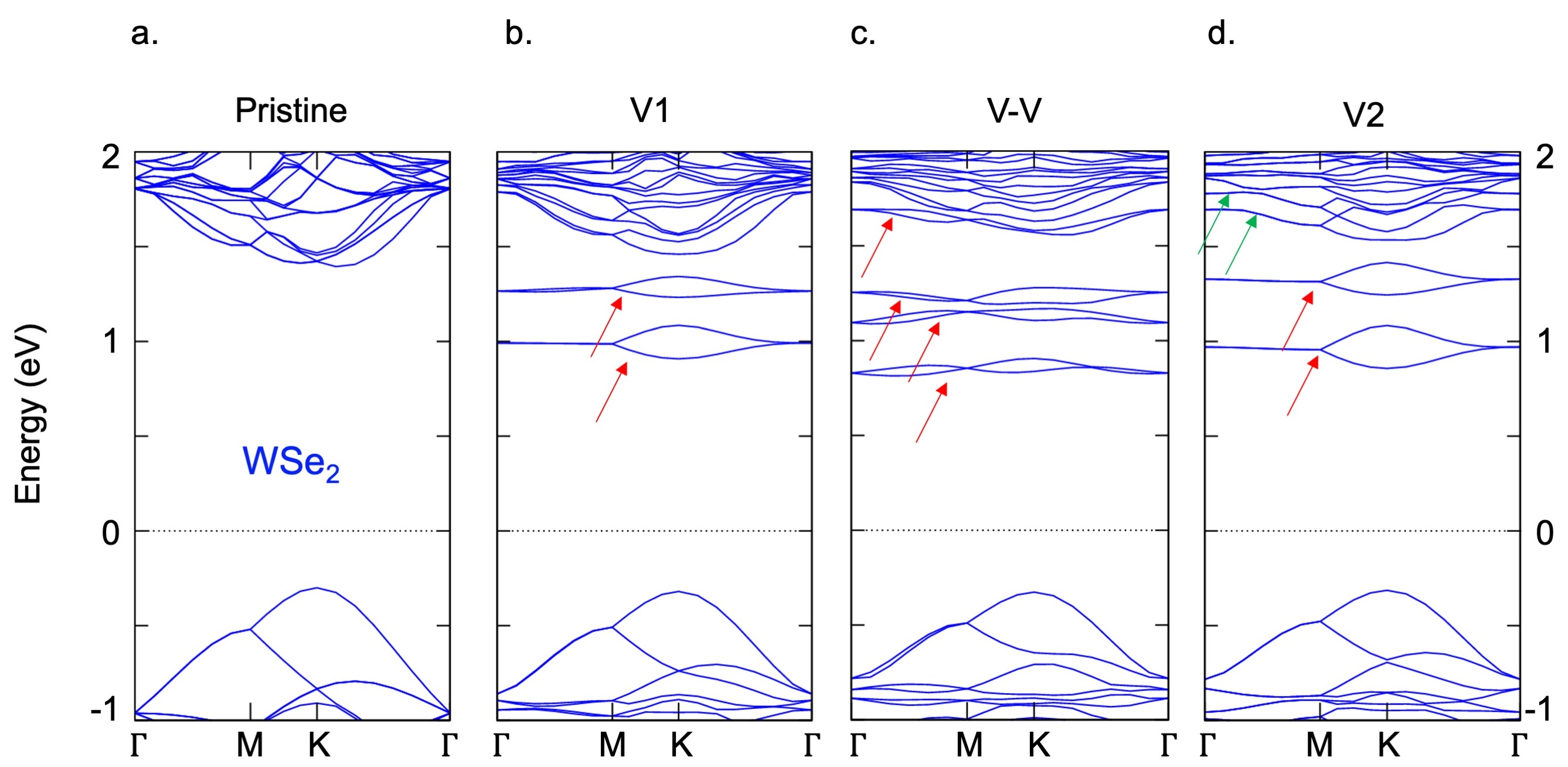}
	\caption{(a) Band structure of pristine $\text{WSe}_{\text{2}}$ showing the characteristic direct band gap at K. (b) Single vacancy band structure showing two defect midgap states, indicated by the red arrows. (c) Lateral divacancy band structure showing four midgap states (red arrows). (d) Vertical divacancy band structure showing two midgap defect states (red arrows) and two additional defect states hybridizing with the conduction band (green arrows).}
	\label{wse2bands}
\end{figure*}

\section{Results}
\subsection{Modeling}

The band structures in this work were modeled by DFT using the Heyd-Scuseria-Ernzerhof (HSE) functional \cite{heyd2005energy}. This high level, range-corrected hybrid functional approximation was chosen to better predict charge localization \cite{moussa2012analysis,khalid2024deep}, leading to more accurate defect state energy levels with respect to the valence band maximum (VBM) and conduction band minimum (CBM) compared to Generalized Gradient Approximation type functionals. We present 6 sets of defect calculations for $\text{WSe}_{\text{2}}$ in addition to the pristine lattice: those of a single top-plane vacancy (V1), lateral top-plane divacancy (V-V), and vertical divacancy (V2) (shown schematically in Figure \ref{vacancyconfigs}) and the same defect geometries with applied strain. In addition to V1 and V2 identified from STEM, we assess the effect of V-V to determine how defect levels and exciton localization differ between laterally spaced divacancies vs. vertically spaced divacancies. As a point of comparison, we also present an identical set for monolayer ($\text{WS}_{\text{2}}$). Although the focus of our experimental results is $\text{WSe}_{\text{2}}$, we present computational results for both $\text{WSe}_{\text{2}}$ and $\text{WS}_{\text{2}}$ to provide a systematic and self-consistent evaluation between materials based on two important parameters individually: defect configuration and chalcogen type.

The HSE band structures for unstrained $\text{WSe}_{\text{2}}$ are shown in Figure \ref{wse2bands}. Pristine monolayer $\text{WSe}_{\text{2}}$ (Figure \ref{wse2bands}a) has a direct band gap between the CBM and VBM. This simple band structure is well-understood in modeling \cite{sun2016indirect, yang2016manipulation}, although experimental spectra exhibit complex emission lines due to the exciton fine structure \cite{shang2015observation}. Direct energy gaps are a critical feature of optical transitions because they facilitate fast rates of radiative decay, indicating their importance not only between band edges for excitonic emission, but between defect states for optical SPE as well. Figures \ref{wse2bands}b-d show the three defect geometries analyzed in this work: V1, V2, and V-V (Figure \ref{vacancyconfigs}). As commonly accepted in the literature \cite{wiktor2016absolute,refaely2018defect}, our calculations show that V1 defects create two midgap states (indicated by red arrows) in the band structure that are degenerate except near the K-point, where a directional dependence emerges for the $x$- and $y$-oriented atomic orbitals \cite{xiang2024charge, kormanyos2015k}. These midgap states are distinct from the pristine band edges for both V1 and V-V, although V-V defects induce four total midgap energy levels. In contrast, the effect of the vertical divacancy V2 is unique. In addition to two non-degenerate midgap states, two more defect states hybridize with and are pushed below the conduction band at K (indicated by the green arrows in Figure \ref{wse2bands}d). This is a critical difference for two primary reasons. First, because excitation of an electron from the VBM directly to a defect state is highly unlikely \cite{hernangomez2023reduced}, defect-bound excitons in TMDs are assumed to emit via defect-mediated relaxation through the midgap states from the conduction band \cite{zhao2013evolution,moody2016exciton}. This explanation has thus far been extended to explain the primary mechanism of SPE. However, from a computational perspective this means that there is no feature that delineates multi-electron defect-bound processes that result in broad, classical emission from a single electron transition resulting in SPE. However, if the conduction band itself becomes hybridized with the defect band, an excitation directly between the VBM and that mixed energy level is more localized to the defect structure. We note that in the projected density of states, V2 defect levels show a relatively higher contribution from selenium p-orbitals and less from tungsten d-orbitals (responsible for delocalized band edge transitions \cite{zhao2013evolution}) compared to V1 and V-V defect levels (Figure \ref{supp-PDOS} in Supplementary Information).

This transition introduces a viable explanation for the energy at which SPE is generally observed in $\text{WSe}_{\text{2}}$ ($\sim$ 1.5 - 1.7 eV), which is higher than the values predicted by relaxation from one or two midgap states and the VBM resulting from a single vacancy ($\sim$ 1.3 - 1.4 eV) \cite{refaely2018defect,wiktor2016absolute}. Although predicting the precise magnitude of a defect-bound transition is difficult in part due to DFT limitations (discussed further in Section \ref{sec:Discussion}), it is pertinent to note that a transition between the VBM and hybrid CBM/defect level predicts a much more accurate energy range of observed SPE, particularly when adjusting for mechanical strain. SI Figure \ref{supp-wse2-strained} shows the same calculations for $\text{WSe}_{\text{2}}$ as in Figure \ref{wse2bands} but with 1$\%$ applied strain. The primary features of the electronic structures resulting from each defect are consistent, except for the separation between the upper and lower levels of each defect state, which generally increases. For V2, this means that a transition between the VBM and the CBM/defect level has a gap magnitude of around 1.7 eV, which is an accurate upper end of the SPE energies observed in the literature \cite{srivastava2015optically, palacios2017large}. Higher degrees of strain would broaden this separation and therefore decrease the energy of the predicted SPE transition, placing the computed energy gap squarely within the observed range and consistent with observations that higher degrees of strain result in lower energy emitters \cite{lu2015atomic,irradiation,kumar2015strain}. With a high enough degree of strain, hybridization between other defect levels can also occur \cite{lu2015atomic,irradiation}, facilitating efficient transitions on the lower energy end of the observed range. The role of strain is complex, and further variables are discussed in Sections \ref{sec:experiment} and \ref{sec:Discussion}.

\setlength{\tabcolsep}{8pt}

\begin{center}
\begin{table}[bh!]
\begin{tabular}{c|c}
	\hline\hline
 \makecell{ Vacancy \\ Configuration} & \makecell{ELF Value} \\ [0.5ex] 
  \hline
    \multicolumn{2}{c}{} \\[-2ex]
    \multicolumn{2}{c}{ \textbf{WSe$_2$}} \\
  \hline
 V1 & 0.46\\ 
 \hline
 V2 & 0.57\\
 \hline
 V-V & 0.34\\
 \hline

 \hline
 \multicolumn{2}{c}{ }\\[-2ex]
\multicolumn{2}{c}{\textbf{WS$_2$} }\\
 \hline
 V1 & 0.61\\
 \hline
 V2 & 0.68\\
 \hline
 V-V & 0.52\\
 \hline

\end{tabular}
\vspace{-0em}
\caption{Electron Localization Function (ELF) of defect configurations. The maximum value of one corresponds to perfect localization of the electron, while the minimum value of zero corresponds to a homogeneous electron gas density.
\vspace{0em}}\vspace{-0em}
\label{table-ELF}
\end{table}
\end{center}

\begin{figure}[h!]
    \centering
    \includegraphics[width=\linewidth]{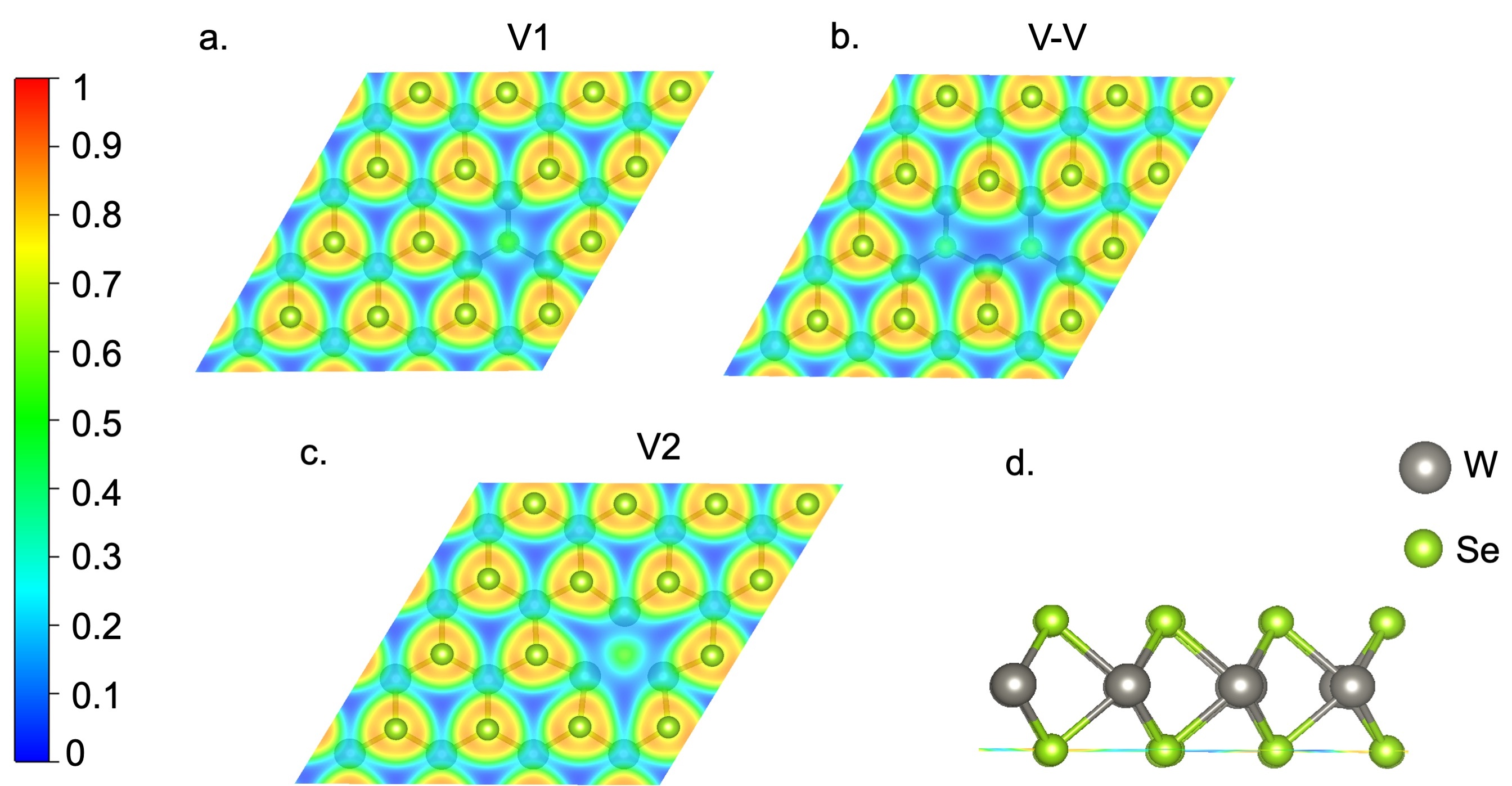}
    \caption{Maps of electron localization for (a) V1, (b) V-V, and(c) V2. For these, localization is determined at the bottom chalcogen plane (d).}
    \label{ELF}
\end{figure}

In addition to the band structures, the electronic localization function (ELF) was calculated for each defect configuration \cite{mandal2024understanding, mandal2025suppressed}. The maps showing the electron density distribution from this function are shown in Figure \ref{ELF}, and the maximum values extracted from the ELF function at the defect sites are tabulated in Table \ref{table-ELF}. Figure \ref{ELF} shows that the vertical divacancy configuration creates a higher degree of localization at the defect site compared to V1 or V-V, with the highest corresponding peak ELF value. These results demonstrate that a vertical divacancy does indeed localize electrons in its vicinity more strongly than single vacancies or neighboring in-plane vacancies. This picture is supported by similar DFT-assisted calculations by Groll~\textit{et al.}, which show charge density distribution modeling of a `selenium column' in 2D $\text{WSe}_{\text{2}}$, which is equivalent  to the V2 vacancy geometry \cite{groll2024dft} shown here. Charge density is localized in the region of the vertical divacancy from orbital hybridization, whereas neighboring lateral vacancies spread out and delocalize electrons, consistent with our results.

\vspace{2em}

\setlength{\tabcolsep}{8pt}

\begin{center}
\begin{table}[bh!]
\begin{tabular}{c|c}
	\hline\hline
 \makecell{ Vacancy \\ Configuration} & \makecell{Formation \\ Energy $E_{\text{f}}$ (eV) } \\ [0.5ex] 
  \hline
    \multicolumn{2}{c}{} \\[-2ex]
    \multicolumn{2}{c}{ \textbf{WSe$_2$}} \\
  \hline
 V1 & 2.82\\ 
 \hline
 V2 & 5.06\\
 \hline
 V-V & 5.51\\
 \hline

 \hline
 \multicolumn{2}{c}{ }\\[-2ex]
\multicolumn{2}{c}{\textbf{WS$_2$} }\\
 \hline
 V1 & 2.93\\
 \hline
 V2 & 5.59\\
 \hline
 V-V & 5.79\\
 \hline

\end{tabular}
\vspace{-0em}
\caption{Formation energies of modeled vacancy configurations.\vspace{0em}}\vspace{-0em}
\label{table-formation}
\end{table}
\end{center}

Finally, the formation energies of each defect geometry presented here within its respective lattice were tabulated, the results of which are shown in Table \ref{table-formation}. Here, we see that V1 has the lowest formation energy alone, which is consistent with its proposed prominence in the literature. However, the next lowest formation energy belongs to V2. Looking specifically at the values for V1 and V2, these results show that vertical divacancies provide greater lattice stability than closely spaced monovacancies in both $\text{WSe}_{\text{2}}$ and $\text{WS}_{\text{2}}$. This is seen by considering twice the formation energy of V1 in each material (2*$\text{V1}_{\text{Se}}$ = 5.64 eV, 2*$\text{V1}_{\text{S}}$ = 5.86~eV) compared to the formation energy of V2 in each  material ($\text{V2}_{\text{Se}}$ = 5.06~eV, $\text{V2}_{\text{S}}$ = 5.59~eV). The formation of two monovacancies is more energetically costly than the formation of one vertical divacancy, and the effect is more pronounced in $\text{WSe}_{\text{2}}$, where the energy difference is -0.58 eV compared to $\text{WS}_{\text{2}}$ with an energy difference of -0.27 eV. This is consistent with the claim that $\text{WSe}_{\text{2}}$ is most likely to have divacancies amongst the TMDs \cite{jeong2019spectroscopic} and highlights the importance of local defect density fluctuations for the formation of SPE as opposed to the average large-scale density. Although an average defect density may be on the order of $10^{12} $cm$^{-2}$, it is a locally higher density that facilitates optimal conditions for the formation of V2, reflected in the high local density of divacancies imaged by STEM.

\begin{figure*}[htbp]
	\centering
	\includegraphics[width=\textwidth]{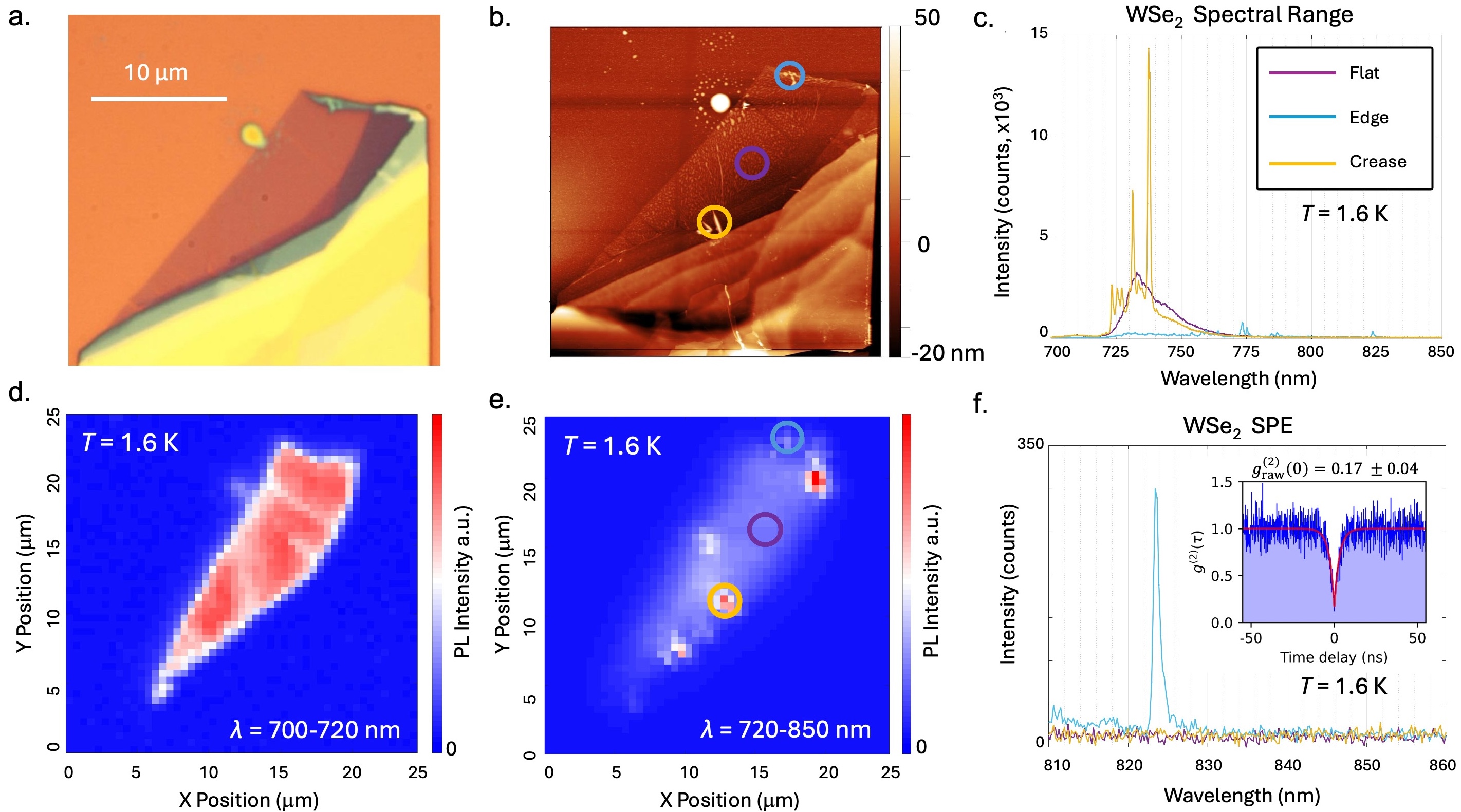}
	\caption{(a) Optical microscope image of $\text{WSe}_{\text{2}}$ monolayer. (b) Atomic force microscopy (AFM) image of the $\text{WSe}_{\text{2}}$ monolayer with areas circled corresponding to smooth (purple), creased (orange), and edge (blue). (c) Low-temperature emission spectra from regions of $\text{WSe}_{\text{2}}$ identified in AFM image. Strained areas exhibit narrow, localized emission features.  (d) Map of monolayer emission integrated over the neutral exciton range (700 – 720 nm), showing the uniform intensity of exciton emission across the flake.  (e) Map of monolayer emission integrated over wavelengths beyond the neutral exciton ($>720$~nm), showing localized, preferential emission near edges, folds, and cracks as identified with AFM. (f) Localized emission feature from the edge strained (blue) spectrum in (c). Inset shows the characteristic antibunching behavior of a single photon emitter in $g^{(2)}(\tau)$ with $g^{(2)}(0) = 0.17$. }
	\label{flatwse2}
\end{figure*}

\subsection{Spectroscopic Properties and the Role of Strain}\label{sec:experiment}

STEM and DFT together demonstrate that the V2 defect configuration is a strong computational candidate to represent a localized single-electron transition in $\text{WSe}_{\text{2}}$. In this section, we provide cryogenic spectroscopic measurements of monolayer $\text{WSe}_{\text{2}}$ prepared by the same method and from the same parent crystal as that used for STEM imaging. Although these measurements do not directly measure the spectral properties of isolated defect structures, they demonstrate SPE in the same material from which the divacancies were identified. This is critical to the assumption that SPE come from naturally occurring defects on which the model is based. One reason why it is so difficult to create a cohesive picture of SPE in the existing literature is the varied parameters with which they are assessed. From paper to paper, computational parameters such as functional and supercell size are different, which alter the outcome of electronic structures. Experimental parameters such as bulk crystal source and the exfoliation method are varied, which can alter the quality of the material, density of the defects, and spectroscopic characteristics. As discussed in the previous section, among related materials in the TMD family, chalcogen type is also not often isolated as a single variable to gauge its effect on exciton localization. Here, both the spectroscopic and photon correlation properties of plain monolayer $\text{WSe}_{\text{2}}$ and $\text{WS}_{\text{2}}$ are presented. The importance of these results for this work is not in a specific novel behavior but rather in the internal self-consistency of the material properties with the other measurements performed. Each monolayer presented here was exfoliated from the same parent crystal using the same methods for all microscopy and spectroscopy measurements, and the experimental conditions were kept constant between materials, thus removing the variables introduced if those parameters were different.
Our work thus far shows that V2 divacancies are experimentally observed in the monolayer lattice, create defect-character transitions in the band structure, are energetically favorable, and localize electrons more strongly than other configurations. If these metrics are effective proxies for quantum emission, our material should host SPE, as is typically the case for strained $\text{WSe}_{\text{2}}$ monolayers. To ensure that our results are self-consistent and do not make assumptions based on literature, photoluminescence (PL) mapping was performed on monolayer $\text{WSe}_{\text{2}}$ at a temperature of $T = 1.6$ K. Consistent with early reports \cite{tonndorf}, neutral exciton emission is shown to be comparable throughout the entire monolayer area (Figure \ref{flatwse2}d), while defect-bound emission is intense at edges and areas of strain (Figure \ref{flatwse2}e) \cite{xu2024sub}. Both strained areas (crease and edge) show sharp, localized emitters that are candidate SPE all the way from 730~nm to 825~nm, as discussed in previous sections. The lowest energy emitter was selected for ease since it is most spectrally isolated from other emission. It is shown in the blue spectrum of (Figure \ref{flatwse2}c,f) and is identified as a SPE source by measurement of the second-order photon correlation $g^{(2)}(\tau)$ (Figure \ref{flatwse2}f, inset). To extract a normalized $g^{(2)}(0)$ value, the data is fit so that the large $\tau$ value corresponding to coherent light is equal to one. Finding that $g^{(2)}(0)< 0.5$ confirms this is a single photon source. Atomic force microscopy identifies strain manifesting as microscopic cracks and wrinkles (Figure \ref{flatwse2}b), with locations of strain corresponding to locations of SPE (Figure \ref{flatwse2}e). SPE is typically observed in strained folds of more than 100 $\text{nm}^{\text{2}}$ \cite{cho2023spatial}, in annular regions around nanopillars of area greater than 100,000 $\text{nm}^{\text{2}}$ \cite{chemomech,peng2020creation,xu2024sub}, or on the shoulders of nanoindentations of about 500~nm in diameter \cite{rosenberger2019quantum,graphite,gavin2025high, lee2025voltage}, so prominent divacancy populations are expected in any typical strained $\text{WSe}_{\text{2}}$ configuration. Further SPE in this sample was identified and characterized (SI Figure \ref{supp-allSPE}). As a point of comparison, the same measurements were performed for monolayer $\text{WS}_{\text{2}}$, found in SI Figure \ref{supp-ws2mono} . Unlike $\text{WSe}_{\text{2}}$, SPE in $\text{WS}_{\text{2}}$ is not observed for three typical monolayers with strain regions. This distinct difference in behavior in our direct comparison of tungsten-based TMDs, implicit in the body of prior literature, highlights how these two materials have different SPE phenomenology. Therefore, the differences between the two tungsten-based TMDs, namely the difference in appropriate defect formation demonstrated by our computational results, should illuminate the fundamental defect origin of SPE from $\text{WSe}_{\text{2}}$. The proposed origins of this distinct behavior are discussed in Section \ref{sec:Discussion}.

\section{Discussion}\label{sec:Discussion}

The abundant work demonstrating quantum emission from $\text{WSe}_{\text{2}}$ highlights the versatility of 2D TMDs for applications in quantum engineering. However, the material platform cannot be fully harnessed until the physical mechanisms of SPE are demonstrated and defined. Targeting the measurement of single defects motivates an abundance of future work; in the mean time, systematic explorations of variables such as defect geometry help fill the gaps in our understanding. Single chalcogen vacancies are commonly identified as the source of SPE across TMDs because they exist in relative abundance and produce midgap energy levels in computed band structures. Although these assumptions are reasonable, they make no distinction between classical defect-bound emission and single electron transitions and also fail to predict an accurate energy range of observed SPE. The candidate electronic structure for optically-active quantum defects, such as nitrogen vacancy centers in diamond, consists of one occupied and one unoccupied defect level within the band gap~\cite{gupta2018two,thomas2024substitutional}. However, theoretical models do not meet this criterion for TMDs and collectively provide competing explanations as to which atomic defects and conditions enable SPE \cite{irradiation,chemomech,cui2018effect,kapuscinski2021rydberg,hernandez2022strain, zheng2019point,zhang2017defect}. Furthermore, the variability of SPE phenomena across the TMD family complicates the identification of a single underlying mechanism that explains the full picture. Given this distinct behavior observed both in the literature and identified here, our work builds a new computational story based on a robust model from defect geometries identified by microscopy, with PL and photon counting included for an internally self-consistent reference. For our purposes, a grounded mechanistic explanation of SPE in $\text{WSe}_{\text{2}}$ should meet three primary requirements: (1) the defects responsible for SPE should be energetically preferred and observed in typical monolayers; (2) the responsible defects should result in direct, localized transitions in the computed band structure; and (3) the results should be self-consistent based on material quality, preparation, and existing experimental results. 

Addressing the first and second requirements, our work isolates \textit{vacancy geometry} as the variable to understand how defect-character transitions form within $\text{WSe}_{\text{2}}$. Experimentally abundant SPE in $\text{WSe}_{\text{2}}$ should arise from a relatively abundant and naturally occurring defect. By imaging vertical divacancies with STEM and comparing their formation energy with the other geometries, it is shown that V2 is present and energetically preferred in the lattice over V-V and closely spaced monovacancies. Additionally, the vertical configuration localizes electrons more effectively than laterally spaced vacancies, as demonstrated by ELF calculations. Within the computed band structures, the key difference between the vacancy configurations is the non-degenerate defect levels mixing with the conduction band as a result of the vertical V2 defect geometry, creating radiative, defect-localized transitions with accurate band gap energies. How these band structures change with supercell size is discussed in SI Section \ref{supp-Appendix:calc}.  Together, this is strong evidence that V2 is a good computational proxy for experimental SPE.

Addressing the third requirement, this work also contextualizes the findings of $\text{WSe}_{\text{2}}$ by isolating \textit{chalcogen type} as a variable to understand the differences in exciton localization between $\text{WSe}_{\text{2}}$ and $\text{WS}_{\text{2}}$. Although V2 geometry also creates a hybrid CB/defect level in $\text{WS}_{\text{2}}$ in our calculations (SI Figure \ref{supp-ws2-bands}), PL and photon counting reveal that SPE is not readily observed in folded areas of a regular monolayer as it is in $\text{WSe}_{\text{2}}$ (SI Figure \ref{supp-ws2mono}). Although SPE has been reported in $\text{WS}_{\text{2}}$, it is always under engineered conditions such as nano-indentation and electrostatic gating \cite{lee2025voltage}, hBN-encapsulated domes \cite{ws2domes}, or chemical functionalization \cite{dasgupta2025carbene}. To the best of our knowledge, no reports exist that show SPE observed from regular $\text{WS}_{\text{2}}$ monolayers, as has been the case for $\text{WSe}_{\text{2}}$ for the last decade. Such a discrepancy in the experimental conditions necessary for SPE between these materials has not yet been explained. In our picture, the foundational mechanism for SPE is present in the $\text{WS}_{\text{2}}$ band structure, but the difference is the prevalence of the necessary defect (V2) in the lattice. Therefore, we postulate that the lack of observed SPE in regular monolayer $\text{WS}_{\text{2}}$ is due to the lack of intrinsic V2 defects, and the observation of SPE in the aforementioned works is engineered by other conditions. This work is also consistent with other existing literature that studies defects in TMDs. It supports the idea that monovacancies in $\text{WSe}_{\text{2}}$ form divacancies more readily than in $\text{WS}_{\text{2}}$ \cite{jeong2019spectroscopic} and that monovacancies result in a broad defect emission rather than a localized SPE \cite{Hossenetal2024}. Our results show that laterally-spaced defects delocalize electrons, meaning a high density of lateral vacancies are not conducive to SPE, but rather broad defect emission. The conclusions here also translate to prior results for molybdenum TMDs. For example, L-band emission in $\text{MoS}_{\text{2}}$ is attributed to sulfur monovacancies \cite{verhagen2020towards}, and SPE in $\text{MoS}_{\text{2}}$ has been experimentally demonstrated only after defect engineering from techniques such as high energy irradiation from helium ions \cite{ionmos2} or ultra-violet photons \cite{uvmos2}. The irradiation is energetic enough to create defects with high formation energies, such as transition metal vacancies, that may not occur in the as-exfoliated material, much like in $\text{WS}_{\text{2}}$. Additionally, existing results for both $\text{MoS}_{\text{2}}$ and $\text{MoSe}_{\text{2}}$ suggest that quantum emission does not rely upon dark spin-forbidden transitions, since the lowest-lying energetic transition in molybdenum TMDs is spin-available \cite{mose2}. The methods here are well-suited for extension to comparing between $\text{MoS}_{\text{2}}$ and $\text{MoSe}_{\text{2}}$ as the role of strain is isolated from that of defect geometry in favor of discovering conditions that are localized and radiative, independent of dark-state hybridization.

The purpose of this work is to present computational and experimental evidence describing the microscopic origins of SPE in $\text{WSe}_{\text{2}}$ and to identify the distinction between the effect of defect geometry and chalcogen type within tungsten-based TMDs. In this context, the defining characteristic is the hybridization of the CBM/defect level from V2 defects and the localized transition it creates. Although TMDs are one category of material with similar properties, each compound is unique and, as such, the interaction of various defects will have different effects on other TMDs. For those, the defect geometry that results in SPE conditions need not be exactly the same. Therefore, further investigation is necessary to understand the origins of SPE in the other TMDs. This work offers a new framework in which to understand quantum emission in $\text{WSe}_{\text{2}}$ and its unique place among the TMDs.

\section*{Author Contributions}

SCG conceptualized the work and performed PL and photon counting. MK executed first principles calculations based on initial development by CJZ. JW performed STEM imaging on samples prepared by JP with technical input from BZ. AD and YL prepared and characterized pristine materials for PL spectroscopy. SCG wrote the manuscript with contributions from all authors. FJH, TJM, MCH, GCS, and NPS supervised the project. 

\section*{Competing Interests}

The authors declare no competing interests.

%\section*{Acknowledgments}
\vspace{2em}

\begin{acknowledgments}
This work was primarily supported by the Center for Molecular Quantum Transduction, an Energy Frontier Research Center funded by the U.S. Department of Energy, Office of Science, Office of Basic Energy Sciences, under Award No. DE-SC0021314. SCG acknowledges current support from the Pritzker School of Molecular Engineering at the University of Chicago and the Advanced Institute for Materials Research at Tohoku University, in collaboration with the Materials Science Division at Argonne National Laboratory, for aide in completing the work. Work performed at the Center for Nanoscale Materials, a U.S. Department of Energy Office of Science User Facility, was supported by the U.S. DOE, Office of Basic Energy Sciences, under Contract No. DE-AC02-06CH11357. FJH acknowledges the U.S. Department of Energy, Office of Science, Basic Energy Sciences, Materials Sciences, and Engineering Division through Argonne National Laboratory under Contract No. DE-AC02-06CH11357. Partial support was also provided for the preparation of pristine materials by the National Science Foundation Materials Research Science and Engineering Center at Northwestern University under Award No. DMR-2308691. SCG gratefully acknowledges Jash Jain for writing Python code to efficiently analyze spectral data.

\end{acknowledgments}

\clearpage

%\appendix

\bibliography{Gavin2026refs}

\end{document}